\documentclass[aps,prl,showpacs,groupedaddress,superscriptaddress,twocolumn]{revtex4}
\usepackage{amsmath}
\usepackage{mathrsfs}
\usepackage{amssymb}
\usepackage[dvips]{graphics,color}
\usepackage{graphicx}
\begin{document}
\title{ Coherent atom-trimer conversion in a repulsive Bose-Einstein condensate}
\author{H. Jing$^{1}$, J. Cheng$^2$, and P. Meystre$^*$}
\affiliation{Department of Physics, The University of Arizona,
Tucson, Arizona 85721, USA\\
$^2$School of Physical Science and Technology, South China
University of Technology, Guangzhou 510640,
 People's Republic of China}
\date{\today}
\begin{abstract}
We show that the use of a generalized atom-molecule dark state
permits the enhanced coherent creation of triatomic molecules in a
repulsive atomic Bose-Einstein condensate, with further
enhancement being possible in the case of heteronuclear trimers
via the constructive interference between two chemical reaction
channels.
\end{abstract}
 \maketitle

The experimental realization of Bose-Einstein condensates (BEC) in
dilute atomic vapors has led to a number of spectacular advances
with implications well past the confines of traditional atomic,
molecular and optical (AMO) physics [1], and with profound
relevance for fields ranging from condensed matter physics to
quantum information science. A development of particular relevance
for the present study is the generation and probing of ultracold
molecular samples and of molecular condensates from atomic BEC,
using magnetic Feshbach resonances (FR) and photoassociation (PA)
[2-6] to control the dynamics of the system. In particular, it has
recently been shown that these two techniques can be combined to
achieve the efficient and stable conversion of atoms into
molecular dimers [3].

The goal of this paper is to demonstrate theoretically that these
techniques can in principle be extended to the generation of
molecular trimers. An important new result is that the creation of
{\em heteronuclear} trimers can be significantly enhanced by the
constructive interference of two quantum channels leading to their
formation from a two-component atomic condensate. The basic idea
is to first create highly excited dimers via a standard FR, and
then to couple them to a bound trimer via PA. A key element of the
scheme is to exploit a coherent population trapping (CPT)
technique to prevent the dimer population from becoming
significant throughout the conversion process. Such a scheme has
previously been proposed for the creation of molecular dimers [3],
and has been theoretically demonstrated to be stable for a broad
range of conditions.

This proposed extension of ``superchemistry'' [6] manipulations
from dimers to trimers exploits the existence of three-body bound
states in ultracold atomic samples as the scattering length for
two-body collisions becomes infinite. As discussed in Ref. [7]
this occurs not only for identical particles, a situation first
considered by Efimov [8-10], but also for two identical and one
different particle. In particular, in the case of two heavy
particles and one light particle these authors found that the
heavy-heavy interaction does not matter, and there is an infinite
number of bound trimer states as long as the light-heavy
scattering length becomes infinite. In the case of the interaction
between one heavy and two light particles, by contrast, there is
no infinite number of bound states unless both two-body scattering
lengths become simultaneously infinite.

We mentioned that a point of particular interest in the formation
of heteronuclear trimers A$_2$B is the role of quantum
interferences between the two paths that involve intermediate
dimers A$_2$ (path AA) and AB (path AB), respectively. To set the
stage we consider first the path AA, whereby AA dimers formed via
a FR are photoassociated with an atom B to form the trimer A$_2$B
[3]. The essence of the idea is to minimize the occupation of
intermediate dimers by exploiting an atom-molecule dark state that
permits the direct association of atoms into trimers without the
formation of a substantial dimer population in the process [5]. We
note that while the proposed scheme is experimentally challenging,
recent progress in the manipulation of dimer-atom resonances
[11-14] indicates that it might become realizable in the near
future. Note also that this process is quite different from a
FR-induced dimer-trimer mixture in a resonant condensate [15].

Our model system consists of a Bose-Einstein condensate of atoms
coupled to molecular dimers via a FR, these dimers being in turn
photoassociated to the atoms to form bound trimers. Denoting the
strength of the atom-dimer coupling with detuning $\delta$ by
$\lambda'_1$, the Rabi frequency of the PA laser by $\Omega'_1$
and its detuning by $\Delta$, the dynamics of the system is
described at the simplest level by the model Hamiltonian
\begin{align}
 \hat{{\cal H}}=&-\hbar\int{{ d} {r}}\biggr\{
 \sum_{i, j}\chi_{i j}\hat{\psi}_{i}^{\dag}(r)\hat{\psi}_{j}^{\dag}(r)
 \hat{\psi}_{j}(r)\hat{\psi}_{i}(r)\nonumber\\
 &+\delta\hat{\psi}_{d}^{\dag}(r)\hat{\psi}_{d}(r)
 +\lambda_1'\bigl[\hat{\psi}
 _{d}^{\dag}(r)\hat{\psi}_{a}(r)\hat{\psi}_{a}(r)+h.c.\bigl]\nonumber\\
&+(\Delta+\delta)\hat{\psi}_{g}^{\dag}(r)\hat{\psi}_g(r)
 -\Omega_1'\bigl[\hat{\psi}_{g}^{\dag}(r)\hat{\psi}_{d}\hat{\psi}_{b}+h.c.\bigl]
\biggr\}. \nonumber
\end{align}
Here, the annihilation operators ${\hat \psi}_i(r)$, where the
indices $i,j=a,b,d,g$ stand for atoms (A and B), dimers and
trimers, satisfy bosonic commutation relations, and the collision
terms proportional to $\chi_{i j}$ describe $s$-wave collisions
between these species. (Trimer formation via the path AB only can
be investigated in a similar fashion.) The nonlinear bound-bound
coupling between dimers and trimers is typically induced by a
narrow-frequency, continuous-wave PA laser, for which the
Franck-Condon factor can be calculated by resonant scattering
theory [2, 11] and may be tuned by pumping methods [16-17].

We assume in the following that the main features of the dynamics
are adequately described by a mean-field analysis,
$\hat{\psi}_{i}\rightarrow\sqrt{n}\psi_{i}$, where $n$ is the
initial atomic density. In this limit, the system is described by
the equations of motion ($\hbar=1$)
\begin{eqnarray}\label{psidot}
\frac{d \psi_{a}}{d t}&=&2in \sum_j \chi_{aj} |\psi_j|^2
\psi_{a}+2i\lambda_1\psi_{d}\psi_{a}^{*},\nonumber\\
\frac{d \psi_{b}}{d t}&=&2in \sum_j \chi_{bj} |\psi_j|^2
\psi_{b}-i\Omega_1
\psi_{d}^{*}\psi_{g},\nonumber\\
\frac{d \psi_{d}}{d t}&=&-(\gamma-i\delta)\psi_{d}+2in\sum_j
\chi_{dj}|\psi_j|^2
\chi_d\psi_{d}+i\lambda_1\psi_{a}^{2}\nonumber\\
&&-i\Omega_1
\psi_{b}^{*}\psi_{g},\\
\frac{d \psi_{g}}{d t}&=&2in\sum_j\chi_{gj}|\psi_j|^2
\psi_g+i(\Delta+\delta)\psi_g-i\Omega_1 \psi_d\psi_b,\nonumber
\end{eqnarray}
where $\lambda_1=\lambda'_1\sqrt{n}$, $\Omega_1=\Omega'_1\sqrt{n}$
and the decay rate $\gamma$ accounts for the loss of untrapped
dimers. To reduce these losses we exploit a CPT technique that
relies on the existence of an approximate atom-molecule dark
state. Such techniques are well known in the case of linear
systems, where they permit the transfer of population from an
initial to a final state via an intermediate state that remains
unpopulated at all times. This is the basis for stimulated Raman
adiabatic passage (STIRAP), which achieves this goal via a
so-called counter-intuitive sequence of pulses [2].

CPT and STIRAP rely explicitly on the validity of the adiabatic
theorem, which applies only to linear systems. Unfortunately, in
the situation at hand two-body collisions render the system
nonlinear, and it is not immediately obvious that STIRAP still
works in that case. This problem has recently been investigated in
Ref.[4], which shows that an approximate adiabatic condition can
still be achieved by linearizing the nonlinear system around the
intended adiabatic evolution. If the eigenfrequencies of the
linearized system are real, the associated ``normal modes'' will
not grow in time, and the system is stable. Hence a system
initially prepared in a CPT state will approximately remain in
that state at all times, although the adiabaticity condition may
be difficult to fulfill at the later stages of the evolution. In
contrast, if the eigenvalues of the linearized problem are
complex, the system is dynamically unstable for some parameter
values [3] under which adiabaticity breaks down.

For the specific case of trimer formation it is easily shown that
under the generalized two-photon resonance condition
    \begin{align}
    \Delta=&-\delta+2(2\chi_{ag}+\chi_{bg}-\chi_{gg})nN_{g,s}\nonumber\\
    &+(4\chi_{aa}-2\chi_{ag}+4\chi_{ab}+\chi_{bb}-\chi_{bg})nN_{a,s}
    \end{align}
eqs.~(\ref{psidot}) admit a steady-state solution with no dimer
population,
\begin{equation}
N_{g,s}=\frac13 \left
(\frac{k(\lambda_i/\Omega_i)^2}{1+k(\lambda_i/\Omega_i)^2} \right
),
\end{equation}
where $i=1$ and $k=4$. (For the path AB, Eq. (3) remains the same,
but with $i=2$ and $k=1$, so that $N_{g,s}^{\rm AB}<N_{g,s}^{\rm
AA}$ for the same external parameters. Note that $N_{d,s}=0$ and
the same two-photon resonance condition holds in both cases.)

This suggests that approximate CPT dynamics such that the dimer
population $N_d$ remains small at all times can be achieved for an
appropriate ``counter-intuitive'' time dependence of the laser
detuning $\Delta$ [3] (see also Ref. [18]). Additionally, it may
be possible to exploit a feedback technique [19] to further
stabilize this process.
\begin{figure}[ht]
\includegraphics[width=8cm]{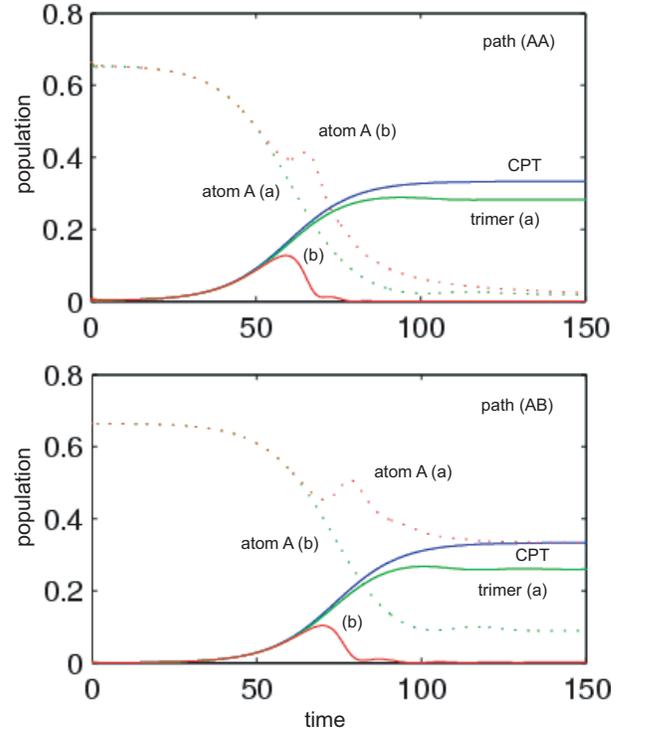}
\caption{(Color online) Heteronuclear trimer population for (a)
$\delta=-3$ (in green) or (b) $\delta=3$ (in red) and their
corresponding numbers of atom A. Time is in units of
$\lambda_l^{-1}$, and $\gamma =1$. The dimer population (A$_2$ or
AB) remains essentially zero at all times and is not shown. The
line labelled "CPT" shows the ideal, analytically derived trimer
population. The evolution of the atoms B is also not shown.}
\end{figure}

Figure 1 shows a numerical simulation of the formation of the
heteronuclear trimers A$_2$B, when considering the channels AA and
AB separately. In this specific example the atom A is $^{23}$Na,
with the $s$-wave scattering length $3.4$nm, and atom B is
$^{87}$Rb, with an $s$-wave scattering length of $5.77$nm,
$\lambda_l$=$4.718\times10^{4}$ ${\rm s}^{-1}$ ($l=1,2$), and
    \begin{equation}
    \label{rabi}
    \Omega_l(t)=\Omega_{l,0}{\rm sech}(t/\tau),
    \end{equation}
with $\Omega_{l,0}/\lambda_l=20$, $\lambda_l\tau$=$20$. The
$s$-wave scattering length for Na-Rb collisions, which depends on
the details of the interatomic potential, is not well known. Here
we take $\chi_{a}$=0.3125, $\chi_{b}$=0.5303, $\chi_{ab}$=0.4214
and the other collisions parameters are taken as 0.0938 (in units
of $\lambda_l/n$) [20]. Very little is known about the scattering
lengths of collisions involving molecular trimers, so we have
carried out simulations with several sets of plausible parameters
[20]. We found that the stable formation of trimers is always
possible for a range of values of the detuning $\delta$.

The curves labelled (a) in Fig.~1 give one such example for
$\delta =-3$. Their general features resemble those of Fig. 2 in
Ref. [4], which corresponds however to the creation of dimers
rather than trimers. In particular, we observe a similarly
increasing departure of the population transfer from the CPT
solution as time evolves. Just as is the case for dimer formation
[3], the association of atoms into trimers is characterized by the
existence of regions with unstable dynamics. This is for instance
the case for for $\delta=3$ (curves (b) in Fig. 1.) As expected,
the two reaction channels lead to different dynamical behaviors.
In particular, the AB channel leaves a significantly larger number
of atoms A in the sample at the end of the conversion process, and
hence results in a lower yield of heteronuclear trimers.

We conclude the discussion of the single-channel cases by noting
that we have also studied the full quantum dynamics of
heteronuclear or homonuclear trimer formation in the short-time
limit using a $c$-number positive-$P$ representation approach [6].
We find that in that limit the quantum dynamics reproduces the CPT
dimer production predicted by the mean-field theory [21], and
quantum noise-induced trimer damping occurs only near a total
atom-trimer conversion [6, 21].
\begin{figure}[ht]
\includegraphics[width=8cm]{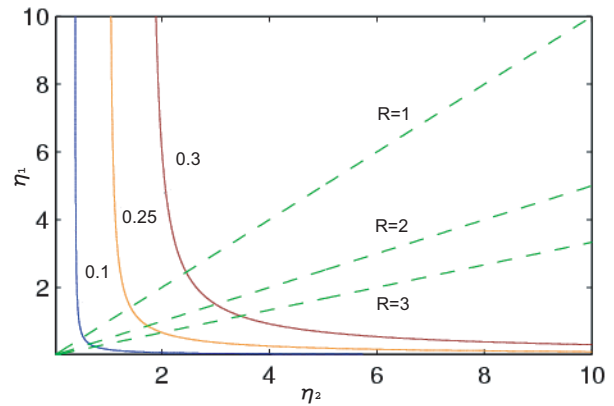} \caption{(Color online)
Normalized CPT trimer population as a function of $\eta_l =
\lambda_l/\Omega_l, l=1,2$. Also shown are three values of $R=
\eta_2/\eta_1$. At constant ratios $\Omega_2(t)/\Omega_1(t)$ the
counterintuitive evolution of the system is along lines of
constant $R$ starting from the origin.}
\end{figure}

\begin{figure}[ht]
\includegraphics[width=8cm]{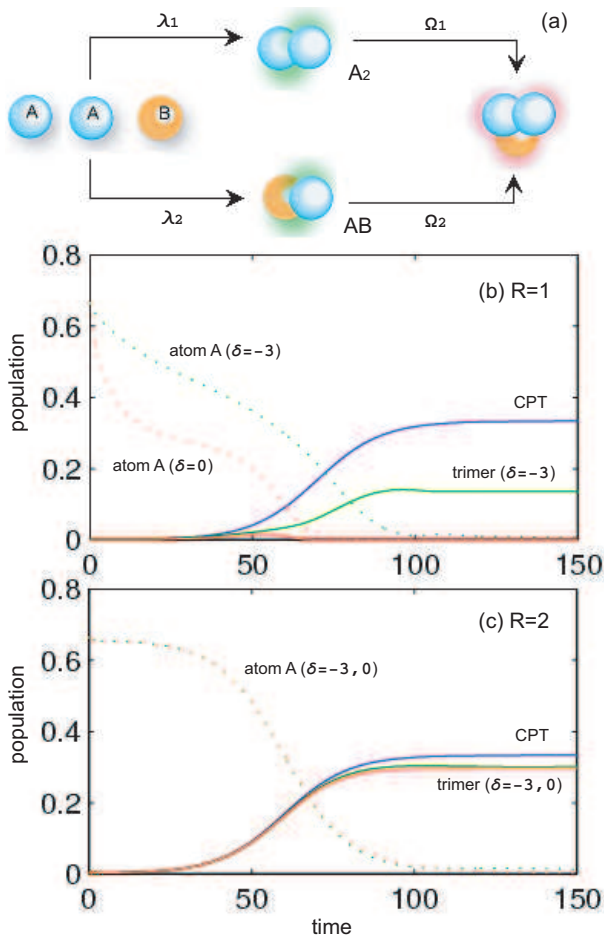} \caption{(Color online)
(a) Two-channel generation of heteronuclear trimers. The time
dependence of the populations of the trimers (solid) and of atom
$A$ (dotted) are shown for $\delta= -3$ and $\delta=0$ and (b)
$R=1$ or (c) $2$ . The CPT value of trimers is also plotted.
$\chi_{d_1d_2}=0$ and $d_{1,2}$ denotes A$_2$ or AB.}
\end{figure}

We now proceed to demonstrate that when acting in concert, the two
channels can yield a significantly larger conversion rate and
approach the ideal CPT yield of 1/3, see Fig. 2. Here, the
coexistence of the two channels provides considerable additional
flexibility in approaching the ideal CPT value for trimer
formation. Note however that this approach either requires an
accidental coincidence of FRs for the A$_2$ and AB dimer
formation, or might be realizable in other cases by applying a
magnetic field gradient [22] across the coexisting A and B
condensates.

The mean-field description of the two-channel situation is a
straightforward extension of the single-channel case. Requiring as
before that the number of dimers remains equal to zero, the CPT
steady-state number of trimers is then
\begin{equation}
N_{g,s}= \frac{(\lambda_1/\Omega_1)(\lambda_2/\Omega_2)^2}
{\lambda_1/\Omega_1+\lambda_2/\Omega_2+3(\lambda_1/\Omega_1)(\lambda_2/\Omega_2)^2},
\end{equation}
where the asymmetry between the two channels results from the fact
that the intermediate dimer involves two undistinguishable
particles in the first one and two distinguishable particles in
the second case.

Figure 2 plots the steady-state trimer number $N_{g,s}$ as a
function of $\eta_l = \lambda_l/\Omega_l, l=1,2$, as well as the
parameter
\begin{equation}
R=\eta_2/\eta_1.
\end{equation}
(Note that there is no CPT solution for $R<0$.) As the STIRAP PA
pulses $\Omega_1(t)$ and $\Omega_2(t)$ are applied, $\eta_1$ and
$\eta_2$ increase and if the ratio of their amplitudes remains
constant the system evolves along a line of constant $R$. To
determine whether an optimum choice of $R$ permits to approach the
ideal trimer population of 1/3 under the non-ideal STIRAP
conditions of our nonlinear system we have solved numerically the
mean-field equations of motion of the system for various values of
$R\in [1,~3]$ (see Fig. 3), using the same parameters as in the
single-channel case.

We found numerically that $R=2$ leads to a trimer production that
most closely approaches the ideal CPT solution, and is
significantly larger than in the single-channel situation of Fig.
1. This is illustrated in Fig.3c for $\delta = 0$ and $\delta =
-3$. Note the insensitivity of trimer production to the detuning
in that case. These results should be contrasted to Fig. 3b, which
shows the evolution of the trimer population in case $R=1$, again
for $\delta =0$ and $\delta = -3$. Here, the trimer production is
very significantly reduced, and depends strongly on the value of
the detuning. Similar results have been obtained for the other
values of $R$ and the other sets of collisions parameters [20]
that we have considered.

We observe also that in the two-channel case the trimer population
can reach a transient value that is larger than its final value.
This suggests that maintaining a constant ratio $R$ during the
evolution of the system may not be optimal. Future work will use
genetic algorithms to determine the optimum time dependence of
$R(t)$ for maximum trimer production.

In conclusion, we have shown that a STIRAP scheme based on
Feshbach-assisted photoassociation, which has previously been
shown to result in the production of ultracold molecular dimers
[3], can be extended to the generation of molecular trimers, and
that in the case of heteronuclear molecules the interference
between two formation channels can lead to a significant
enhancement of trimer production. Future work will study further
ways to optimize this scheme, and will give a quantum-mechanical
description of the system, in particular with the goal of
understanding the role of quantum fluctuations in the early stages
of trimer production and the quantum statistics of the trimer
field. We also plan to study the stability properties and the
adiabatic geometric phase of the system [4], the superchemistry
reaction A$_2+$B $\rightarrow$ AB+B via a dark state technique,
and possibly trimer creation or amplification in an optical
lattice [23]. While experiments along the lines of this analysis
promise to be challenging, recent progress in Efimov
superchemistry [9-12, 15] and in the manipulation of dimer-atom
resonances [11-14] indicates that they may become possible in the
not too distant future.

\acknowledgements

This research is supported in part by the US Office of Naval
Research, by the US National Science Foundation, by the US Army
Research Office, by the Joint Services Optics Program, by the
National Aeronautics and Space Administration and by the NSC and
SCUT. The authors thank Dr. M. Bhattacharya for helpful
discussions, and Dr. P. Julienne for bringing Ref. [7] to their
attention.

\noindent
* Electronic address: pierre.meystre@optics.arizona.edu


\begin{thebibliography}{99}
\bibitem{20} For example, in the Thomas-Fermi approximation, we can
take Na-Rb scattering length as half the critical value $[(M_{\rm
Na}M_{\rm Rb})^{1/2}/(M_{\rm Na}+M_{\rm Rb})](a_{\rm Na}a_{\rm
Rb})^{1/2}$ for the stable coexistence of Na and Rb BECs. Here
$M_{\rm {Na,Rb}}$ and $a_{\rm{Na,Rb}}$ denote the mass and
scattering length of Na and Rb atoms, respectively. Thus, by using
$\chi_{ab}$=0.0832 and assuming the other collisions parameters in
both cases to be those of atom A, except for
$\chi_{\sigma\varsigma}$=-1.4583 ($\sigma$=$a$, $b$ and
$\varsigma$=$d$, $g$), we can reach a similar result as Fig. 1
(with fewer numbers of trimers).
\bibitem{21} Details of the derivations will be published
elsewhere.
\end{thebibliography}
\end{document}